\documentclass[aps,twocolumn]{revtex4}

\usepackage{epsfig}

\begin{document}

\title{Contact process with long-range interactions: a study in the
ensemble of constant particle number}

\author{Carlos E. Fiore} 

\email{fiore@if.usp.br} 

\author{M\'{a}rio J. de Oliveira}

\email{oliveira@if.usp.br} 

\affiliation{Instituto de F\'{\i}sica,
Universidade de S\~{a}o Paulo, \\
Caixa Postal 66318\\
05315-970 S\~{a}o Paulo, S\~{a}o Paulo, Brazil}

\date{\today}

\begin{abstract}

We analyze the properties of the contact process with long-range
interactions by the use of a kinetic ensemble in which the 
total number of particles is strictly conserved. 
In this ensemble, both annihilation and creation processes
are replaced by an unique process in which 
a particle of the system chosen at random leaves its place and jumps to
an active site. The present approach is particularly
useful for determining the transition point and 
the nature of the transition, whether continuous 
or discontinuous, by evaluating the fractal dimension
 of the cluster at the emergence of the phase 
transition. 
We also present another criterion  appropriate to 
 identify the phase transition that 
consists of studying the system in the supercritical regime, where 
the presence of a ``loop'' characterizes the first-order transition.
All results obtained by the present approach are in full
agreement with those obtained  by using the constant rate ensemble,
supporting that, in the thermodynamic limit the results from distinct
ensembles are equivalent.

PACS numbers: 05.70.Ln, 05.50.+q, 05.65.+b

\end{abstract}

\maketitle

\section{Introduction}

The study of equivalence of ensembles
that describe systems in thermodynamic equilibrium,
the Gibbs ensembles,
has a long tradition in statistical mechanics.
Thermodynamic properties evaluated in distinct ensembles
will be the same if they are equivalent. 
In the case of equilibrium systems, the ensembles are set up according
 to well known prescriptions that take account the existence of a
 Hamiltonian \cite{land58,ruel69}.
The possibility of representing nonequilibrium systems in distinct 
ensembles was first
considered by Ziff and Brosillow \cite{ziff92} when they used the
constant coverage ensemble to study first-order transitions in a
catalytic reaction model \cite{ziff86}, originally defined in the
ensemble of constant rate.  Afterwards, Tom\'e and de Oliveira
\cite{tome01} introduced the conserved contact process, a version of
the contact process \cite{marr99,harr74} in the ensemble of constant
particle number. The equivalence between the constant rate and
constant particle number ensembles for the contact process
 in the stationary regime  has been
proved by Hilhorst and van Wijland in the thermodynamic limit
\cite{hilh02}. Later on, the
equivalence of ensembles in the stationary regime 
was extended for other systems with short-range interactions with
distinct annihilation dynamics \cite{oliv03,fior05}.

In equilibrium statistical mechanics, the equivalence of ensembles is granted
for homogeneous and extensive systems  in the thermodynamic limit 
\cite{land58,ruel69,gross}. 
On the other hand, for non-extensive systems, the canonical and
 microcanonical ensembles are not equivalent \cite{bell68}. 
Examples of  physical problems in which canonical and microcanonical
ensembles are not equivalent are
 nuclei, atomic clusters in which the range of forces between their
constituents are comparable to the system size, systems at the 
first-order transition with phase separation \cite{gross,thirring}
and models with long-range forces \cite{gross,barr01,dauxois}. 
In particular, this latter  set of problems  has  
deserved a great interest not only in statistical mechanics, but also
in other areas, such as nuclear physics, astrophysics \cite{bell68} 
and plasma physics \cite{smith}.
Nonequilibrium systems with long-range interactions have also been 
proposed and in this case, 
we may ask whether the equivalence of ensembles may be valid
for such systems.

In the context of nonequilibrium statistical mechanics, systems
displaying long-range interactions
have been proposed originally as more realistic models describing
the   spreading processes,
instead of  short-range systems. In particular, an 
anomalous model of the directed percolation has been 
proposed by Mollison \cite{mollison}. In this problem,
the infection probability  performs a  {\it L\'evy flight}
  decaying with the distance $r$ as a power-law 
relation  $1/r^{\sigma+d}$, where $d$ is the spatial dimension of the
system and $\sigma$ is  control parameter.
The critical behavior of anomalous directed percolation 
displays a  family of universality classes that  
 have been studied  by Grassberger \cite{grass86}, by  Janssen
et al \cite{jan99} who have considered  field-theoretic
renormalization group,  by Hinrichsen and Howard \cite{hinri99}
who performed numerical simulations and recently by Tessone et al 
\cite{tessone} who analyzed  a class of spatially extended chaotic
systems with power-law decaying interactions. In all cases above,
the critical exponents vary continuously with
the parameter $\sigma$.  More recently, 
another class of nonequilibrium systems with long-range interactions,
named contact processes with long-range interactions, 
have been  introduced by Ginelli et al. \cite{gine05,gine06,hinri07}, 
 inspired by  pinning-depinning
transitions in nonequilibrium wetting phenomena.
 By varying the
control parameters, the contact process with long-range interactions
exhibits a richer phase diagram than the usual short-range contact process, 
with distinct universality classes 
\cite{gine06} and discontinuous phase transitions
\cite{gine05,gine06}.

Two aims  concern us in this paper. First,
we wish to analyze the  equivalence of ensembles 
in nonequilibrium systems with long-range interactions.
The equivalence of ensembles 
 has been studied in systems with short-range interactions 
\cite{tome01,saba02,fior05,fior04},
{\it but not} in systems with long-range interactions.
We will consider here the $\sigma-$contact process with long-range
interactions, introduced by Ginelli et al \cite{gine05}.
Second,  we wish to show that the present approach offers a new procedure
for analyzing first-order transitions of systems with absorbing
states.
Due to the existence of the absorbing states,  
standard procedures used successfully in the study of 
discontinuous equilibrium phase transitions may not work well when applied to
describe nonequilibrium first-order transitions. In the present case,
the first evidence for a  discontinuous transition will be given
by the presence of ``loops''  in the rate versus density curve,
as long as the system is finite. As it will be shown, the
loops disappear in the thermodynamic limit giving rise
to a tie line, the true signature of a first-order transition.
Another advantage of using the present approach is verified 
when one analyzes the system at the emergence of  the phase transition, since
 the nature of the phase transition can be inferred simply by 
examinating the structure of particles. As we shall see,
if, at the emergence of the transition, the cluster is
fractal, the transition will be continuous; if the cluster is
compact, the transition will be  discontinuous.
A compact cluster provides us an evidence of a discontinuous transition.

\section{Model}

\subsection{Constant rate ensemble}

The one dimensional $\sigma-$contact process with long-range interactions
\cite{gine05} is defined in a chain of $L$ sites with periodic
boundary conditions as follows. To each site $i$ of a one-dimensional
lattice is attached an occupation variable $\eta_i$ that takes
the values 0 or 1 according whether the site $i$ is empty or
occupied by a particle, respectively. The process is composed
of  spontaneous annihilation of a single particle ($1\to 0$)
and   catalytic creation of a particle ($0\to 1$). Particles
 are created only in empty sites in which
 at least one of its nearest neighbor sites is occupied by a particle.
These empty sites are  named { \it active sites}.
The rate of creation 
depends on the length of the island of empty sites next
to the active site and decreases algebraically with the
size of the island, introducing an effective long-range interaction.  
The total transition rate $w_i(\eta)$ is given by
\begin{equation}
w_i(\eta) = \omega_i^c(\eta)+\alpha \omega_i^a(\eta),
\label{eq1}
\end{equation}
where $\alpha$ is a parameter and
\begin{equation}
\omega_i^a(\eta) = \eta_i,
\label{eq2}
\end{equation}
describes a spontaneous annihilation and
\[
\omega_i^c(\eta) 
=\frac12\, 
\sum_{\ell=1}^\infty  (1+\frac{a}{\ell^\sigma})\eta_{i-1} {\bar\eta}_i
{\bar\eta}_{i+1}\ldots{\bar\eta}_{i+\ell-1}\eta_{i+\ell}
\]
\begin{equation}
+\frac12\,  
\sum_{\ell=1}^\infty (1+\frac{a}{\ell^\sigma})\eta_{i+1}{\bar\eta}_i 
{\bar\eta}_{i-1}\ldots{\bar\eta}_{i-\ell+1}\eta_{i-\ell},
\label{eq3}
\end{equation}
describes a catalytic creation that depends on the length $\ell$
of the island of empty sites, where $a$ and $\sigma$ are
parameters and  we have considered the shorthand notation 
${\bar\eta}_i\equiv 1-\eta_i$ in Eq. (\ref{eq3}).
When $a=0$ one recovers the original short-range
contact process \cite{marr99,harr74}.

Here, numerical simulations in the constant rate
ensemble is performed 
as follows.  A particle is chosen at random from a list of occupied sites.
It is annihilated with probability $p=\alpha/(1+a+\alpha)$. With
probability $1-p$, a new particle may be created next to the chosen
particle. This is done by choosing first one of its two nearest
neighbors  with equal probability
and then a particle will be actually created with probability
 $q=(1+a\ell^{-\sigma})/(1+a)$ provided the chosen site is empty.  
This algorithm gives the following
ratio between the creation  and annihilation  
of  particles 
$(1-p)q/2p=(1+a\ell^{-\sigma})/2\alpha$, which is equivalent to that
considered by Ginelli et al \cite{gine05}, in which 
 the creation and annihilation of particles
occur with rates $\lambda(1+a\ell^{-\sigma})$ and $1$, respectively as
long as $\alpha$ is related to $\lambda$ by $\alpha=1/2\lambda$. 
The increment of the time is given by $1/N_p$, where $N_p$ is the
 number of occupied sites.

For large values of $\alpha$, the system is constrained into the
absorbing state, in which no particles are allowed to be created.
Decreasing the parameter $\alpha$, a phase transition to 
an active state takes place, whose  location  depends
on the parameters $a$ and $\sigma$. For a fixed value of $a$
(we take here to be $a=2$, as considered by Ginelli et al.
\cite{gine05}) and $\sigma>1$ the transition is continuous with
critical exponents belonging to direct percolation (DP)
universality class. For $0<\sigma<1$ the transition 
becomes discontinuous.
Ginelli et al \cite{gine05} found that, 
the crossover between the discontinuous  and second-order phase
transitions  occurs at $\sigma=1.0$.
\subsection{Constant particle number ensemble}

In the constant particle number ensemble, 
the control parameter is the total particle number $n$.
Again we used a chain with $L$ sites with periodic boundary conditions.
 Particles are neither created nor annihilated. Instead, a particle leaves
their place and jump to an empty site. However, the jump process is
not unrestrictive process, since their occurrence must be consistent
with the rules of their version in the constant rate ensemble. 
More specifically, the dynamics 
that characterizes this ensemble  is defined as follows.
A particle of the system, chosen at random,
 leaves its place, located for example at the site $i$, and  
jumps to an empty site placed at the site $j$, also chosen at random.
The jump rate will depend on the specific
rule of the considered model.
One may define the jump process by a dynamics
in which  both creation and annihilation occurs simultaneously, 
according to the transition rate 
$w_{ij}(\eta)$  is given by \cite{tome01,oliv03,fior05}
\begin{equation}
w_{ij}(\eta) = \omega_i^a(\eta)\omega_j^c(\eta).
\label{eq4}
\end{equation}
To see that this transition rate leads to a dynamic that is equivalent
to that given by Eq. (\ref{eq1})  let us consider the
total rate $\sum_i w_{ij}(\eta)/L=\sum_i\omega_i^a(\eta)\omega_j^c(\eta)/L$ in which particles jump to site $j$. 
In the thermodynamic limit the system size $L\rightarrow \infty$ and the 
sum $\sum_i \omega_i^a(\eta)/L$ approaches $\langle\omega_i^a(\eta)\rangle$, 
by the law of large numbers,  
so that $\sum_i w_{ij}(\eta)/L = \langle\omega_i^a(\eta)\rangle 
\omega_j^c(\eta)$.
By an analogous argument the total rate in which particles
leave the site $i$ is
$\sum_j w_{ij}(\eta)/L = \langle \omega_j^c(\eta)\rangle \omega_i^a(\eta)$.
Comparing with the Eq. (\ref{eq1}), the averages
 $\langle \omega_i^a(\eta)\rangle$ and 
$\langle \omega_i^c(\eta)\rangle$ should be proportional to $1$ and
$\alpha$, respectively, thus 
\begin{equation}
\label{eq5}
{\bar\alpha} = \frac{\langle\omega_j^c(\eta)\rangle}
{\langle\omega_i^a(\eta)\rangle},
\end{equation}
where $\omega_j^c(\eta)$ and $\omega_i^a(\eta)$ are given by Eqs. 
(\ref{eq2}) and (\ref{eq3}), respectively, so that Eq. (\ref{eq5})
is given by
\begin{equation}
\label{eq5}
{\bar\alpha} = \frac{1}{\rho}\langle\omega_j^c(\eta)\rangle,
\end{equation}
where $ \langle\omega_i^a(\eta)\rangle= \langle \eta_i \rangle=\rho $ 
is the density of particles.
This formula allow us to evaluate the parameter ${\bar\alpha}$
within respect to the ensemble of constant particle number.
The average
$\langle\omega_j^c(\eta)\rangle$  can be understood 
 as  ``density of active sites'', 
that is, the density of empty sites in which particles may jump to them.
It  can be evaluated directly from numerical simulations by
computing Eq. (\ref{eq3}) that is zero and
$(1+a\ell^{-\sigma})/2$ according  whether the site $i$ is
occupied or empty, respectively, where $\ell$ is the
length of the island of inactive sites in which the active site
is located.
For time-dependent regime, however, Eq. (\ref{eq5}) might not always
hold. For instance, if the initial state is such that averages $\langle\omega_j^c(\eta)\rangle$ and
$\langle\omega_i^a(\eta)\rangle$ are not constants, Eq. (\ref{eq5}) cannot be satisfied \cite{hilh02,oliv03}.

The actual numerical simulation of the constant particle number
ensemble is realized as follows.
An empty site surrounded by at least one particle (active site) 
is  chosen at random. Next, a particle of the system, also chosen
at random, jumps to the active site with probability 
$p_\ell=(1+a\ell^{-\sigma})/(1+a)$.
The constant factor $1/(1+a)$ is used in order to guarantee
 that $p_\ell\leq 1$, since $1+a\ell^{-\sigma}$ may be greater
than $1$. In the constant rate ensemble, the control parameter may be included
in the probability of choosing each subprocess (creation or
annihilation).
However,  in the present case, as both processes occur simultaneously
and the parameter $\bar{\alpha}$ is not constant, this procedure is
not possible. 

In contrast to the constant rate ensemble, the conserved
contact process does not have, strictly speaking, an absorbing state.
This fact constitutes an useful tool in the study of
phase transitions because there is no danger
of falling into the absorbing state as happens in numerical
simulations of the constant rate ensemble.
In the conserved ensemble (constant particle number ensemble) 
the equivalent of the
absorbing state is the state with zero density, as
is the case of an infinite system with a finite number
of particles, also named subcritical regime.

\section{Cluster approximations}
The first analysis we have performed here consists of studying the 
$\sigma-$contact process by means of cluster approximations.
We will consider here approximations at the  level of one and two sites.

At the level of one-site approximation, we use the following
approximation for the probability of a string of sites:
\begin{equation}
P(\eta_1,\eta_2,...,\eta_\ell)=P(\eta_1)P(\eta_2)...P(\eta_\ell).
\end{equation}
In this case, it suffices to consider
the dynamic variables  $P(1)=\rho$ and $P(0)=1-\rho$. 
In the steady state, one has the following relation 
\begin{equation}
\alpha=1-\rho+a\rho\sum_{\ell=1}^{\infty}\frac{(1-\rho)^{\ell}}{\ell^\sigma}.
\label{eqmf}
\end{equation}
This equation has already been obtained by Ginelli et al
\cite{gine05,hinri07}.
For $0<\sigma<1$, Eq. (\ref{eqmf}) establishes a discontinuous
transition between an absorbing and an active  state that can be
viewed by the existence of ``loops'' \cite{loops}.  
As it will be
seen later, numerical simulations in the constant particle number
ensemble  also presents ``loops'' for  finite
systems in the interval $0<\sigma<1$.
However, Eq. (\ref{eqmf}) also exhibits loops for  $\sigma>1$, in 
contrast to  numerical simulations. 

\begin{figure}
\epsfig{file=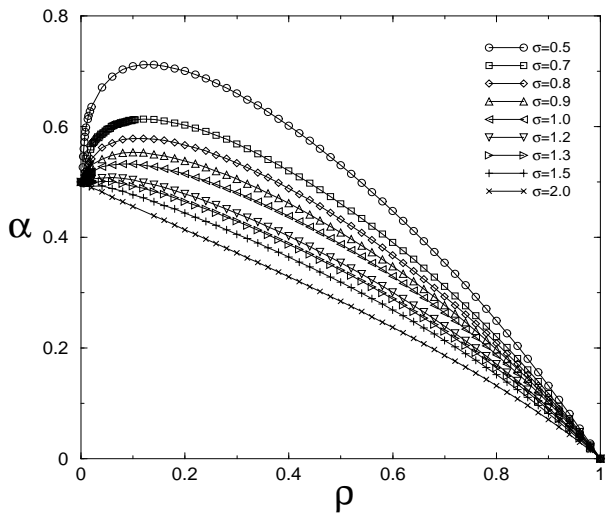,width=8cm,height=7cm}
\caption{Quantity $\alpha$ versus $\rho$ for several values
of $\sigma$ at the level of two-site mean field approximation. 
A crossover from a
first-order to a continuous transitions occurs at $\sigma=(1.3 \pm
0.1)$. Symbols have been used in order to distinguish the distinct
values of $\sigma$.}
\label{fig1a}	
\end{figure}

In order to obtain improved results, we have derived relations by
considering correlations of two sites.
In this case, the probability of a string of sites is approximated by 
\begin{equation}
P(\eta_1,\eta_2,\eta_3,...,\eta_\ell)=
\frac{P(\eta_1,\eta_2)P(\eta_2,\eta_3)...P(\eta_{\ell-1},\eta_{\ell})}
{P(\eta_2)P(\eta_3)...P(\eta_{\ell-1})}.
\end{equation}
The dynamic variables are  now $P(1),P(0)$ and   
nearest-neighbor joint probabilities $P(11),P(10)=P(01)$ and $P(00)$.
Only two of them are independent,
so that it is necessary to write down two equations. By solving these
equations numerically, we find  the behavior of   
 $\alpha$ versus $\rho$ for several values of $\sigma$, as showed in
Fig \ref{fig1a}.  
 In contrast to the previous case, two-site approximations
supports a change in the nature of the phase transition by increasing
the parameter $\sigma$, passing from first-order to  second-order 
at $\sigma=(1.3 \pm 0.1)$.
This can be identified by  disappearance of loops. Although 
the results obtained from two-site  approximations  agree
qualitatively with numerical simulations, the location of the 
transition points as well as the value of  $\sigma$ that characterizes 
the crossover between first-order and second-order transitions  are incorrect.

\section{Numerical results}
 Numerical simulations of the $\sigma-$contact process were
performed   for   $a=2$ and several values
of  $\sigma$.
We have used $2\times10^7$ Monte Carlo steps to evaluate
the appropriate quantities, after discarding a sufficient
number of steps to reach the stationary state. Here one Monte Carlo
step corresponds to $n$ jumping processes.
In order to compare results obtained from distinct ensembles
we have simulated the contact process with long-range
interactions not only in the constant particle number
ensemble, but also in the constant rate ensemble.
We have first determined the quantity ${\bar\alpha}$
in the conserved (constant particle number) ensemble, by using Eq. (\ref{eq1}),
for several densities $\rho$. Next we used the values of ${\bar\alpha}=\alpha$
to perform numerical simulations in the constant
rate ensemble which in turn gives us the average density ${\bar\rho}$. 
According to Table I, the excellent agreement between the results
 confirms the equivalence of ensembles.
However, numerical simulations  provide distinct
results  at the phase coexistence, even when one considers  
large system sizes.  For example, for $L=20000$ and $\sigma=0.5$, 
simulations in the constant rate ensemble for $\alpha=0.42313$ gives
$\bar{\rho}=0.76600(4)$.
This value of $\alpha$  corresponds
to two densities in the constant particle number ensemble 
  $\rho=0.60$ and $\rho=0.766$. As it will be shown later, 
both ensembles become equivalent at the phase coexistence in the
thermodynamic limit.

From now on we will drop the bars over ${\bar\alpha}$ and ${\bar\rho}$.
\begin{table}
\begin{center}
\caption{Results of numerical simulations coming from
the constant particle number ensemble (second and third
columns) and from the constant rate ensemble (fourth and fifth 
columns) for a  system  size $L=20000$.}
\begin{tabular}{ccccc}
\hline
$\sigma$ & $\rho$ & ${\bar\alpha}$ & $\alpha$ & ${\bar\rho}$ \\
\hline
2.0  &  0.7500  &  0.2964(4)   &  0.29638  &  0.7509(9)  \\
1.5  &  0.6900  &  0.33481(1)  &  0.33480  &  0.6900(1)  \\
1.0  &  0.7000  &  0.37007(1)  &  0.37000  &  0.7004(2)  \\
0.5  &  0.8500  &  0.32887(8)  &  0.32890  &  0.85008(8) \\
0.4  &  0.9000  &  0.24948(2)  &  0.24948  &  0.90000(1) \\
\hline
\hline
\end{tabular}
\end{center}
\end{table}

\subsection{Subcritical regime}
\begin{figure}
\epsfig{file=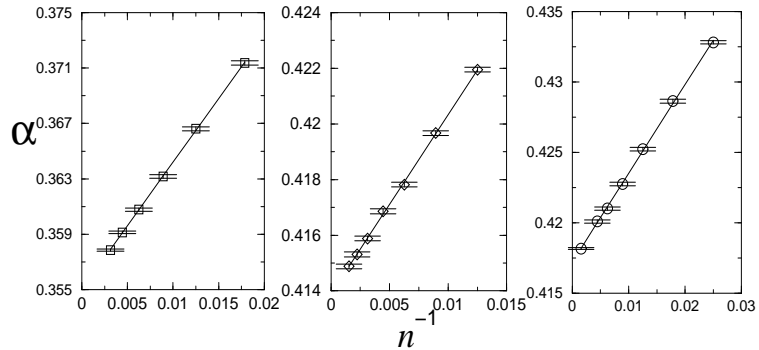,width=10cm,height=5cm}
\caption{Quantity $\alpha$ versus $1/n$ for the
subcritical regime. The left, center and right panels
correspond to $\sigma=2$, $\sigma=0.5$ and $\sigma=0.4$, respectively.
The straight lines fitted to the data points give the
extrapolated critical points $\alpha_0=0.35497$, 
$\alpha_0=0.41389$ and $\alpha_0=0.41714$.}
\label{fig1}	
\end{figure}

Let us consider a finite number of particles placed on
infinite lattice. In this situation the density of particles vanishes
and the system is constrained to remain in the subcritical
regime. 
The actual simulation of the subcritical regime is done by using a 
finite lattice and check whether a particle
reaches the border. 
If a particle
never reaches the border, the system size may be taken as infinite.
The size of the system was chosen to be big enough, so that no
particles have reached the border for the maximum time considered.
An important feature of the subcritical regime
is that the increase of the total number of particles 
makes the system approach the transition point at $\alpha=
\alpha_0$. 
The expected value of $\alpha$ obtained for a fixed number of 
particles $n$ approaches its asymptotic value $\alpha_0$
according to \cite{tome01,saba02,fior05,fior04}  
\begin{equation}
\label{eq6}
\alpha-\alpha_0 \sim \frac{1}{n}.
\end{equation}
A linear extrapolation of $\alpha$ versus $1/n$ when $n\to\infty$
gives $\alpha_0$. In Fig. \ref{fig1} we have plotted the numerical
values of $\alpha$ determined from simulations of an infinite
system with $n$ particles for different values of $\sigma$.
Numerical extrapolations obtained by using Eq. (\ref{eq6}) give us
$\alpha_0=0.41714(6)$, $\alpha_0=0.41389(1)$ and 
$\alpha_0=0.35497(6)$ for $\sigma=0.4$, $\sigma=0.5$ and 
 $\sigma=2$, respectively, which  agree very well with  estimates
 $\alpha_0=0.4172(1)$, $\alpha_0=0.41382(3)$ and $\alpha_0=0.3548(2)$,
obtained from the constant rate ensemble. The
same agreement is verified for other values of $\sigma$,
revealing the utility of the present procedure for locating the transition
point.
In the section IV C, we will show   results 
from time-dependent numerical simulations for $\sigma=0.4$.
For $a=0$, the phase transition takes place at $\alpha_0=0.303228(2)$,
as expected for the usual short-range contact process \cite{marr99}.
\begin{figure}
\centering
\epsfig{file=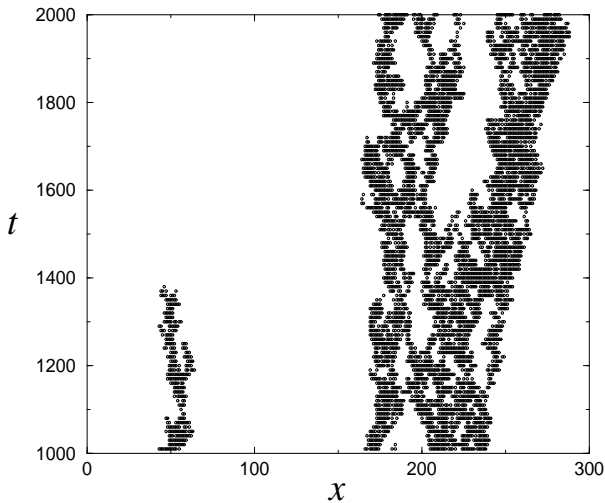,width=8cm,height=7cm}
\caption{Typical run for the conserved long-range contact
process with $n=50$ particles for $\sigma=2$ starting from
a random configuration. A unit of time, or a Monte Carlo step,
corresponds to $n$ particle jumps. We have discarded
$3\times 10^{6}$ initial Monte Carlo steps.}
\label{fig2}
\end{figure}

\begin{figure}
\centering
\epsfig{file=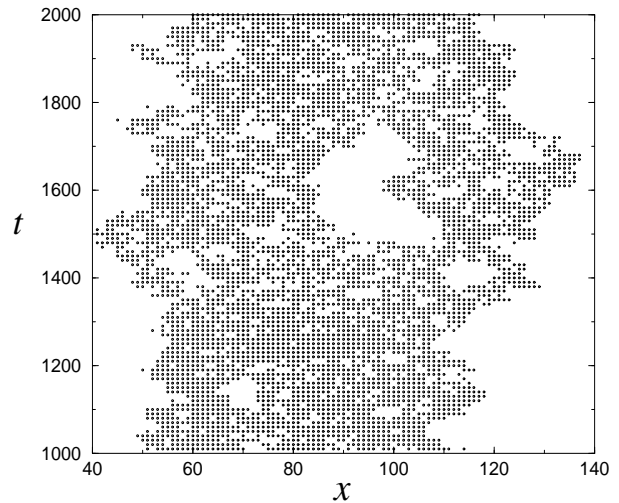,width=8cm,height=7cm}
\caption{Typical run for the conserved long-range contact
process with $n=50$ particles for $\sigma=0.5$ starting from
a random configuration. A unit of time, or a Monte Carlo step,
corresponds to $n$ particle jumps. We have discarded
$3\times 10^{6}$ initial Monte Carlo steps.}
\label{fig3}
\end{figure}
 The first procedure considered here for classifying  
the phase transition consists of examinating the
spatial structure of particles at the emergence of the phase
transition.  In Figs. \ref{fig2} and \ref{fig3} we have plotted
 the time evolution
for two distinct values of $\sigma$. For $\sigma=2$, the system generates
 fractal clusters, whereas for  $\sigma=0.5$ they are compact.

A measure of the size of the cluster of particles is 
given by the quantity $R$ given by
maximum distance between two particles of the cluster
\cite{saba02,brok99}.
 As long as $n$ is finite, $R$ is also finite, 
but it diverges when $n\rightarrow 
\infty$. It is related to the  total particle number $n$ through the 
relation
\begin{equation}
R \sim n^{1/d_F},
\label{fractal}
\end{equation}
where  $d_F$ is the fractal dimension \cite{saba02,brok99}. The 
quantity $R$ will be evaluated here by measuring the end-to-end spread
 of the cluster \cite{saba02,fior05,fior04,brok99}. 
\begin{figure}
\epsfig{file=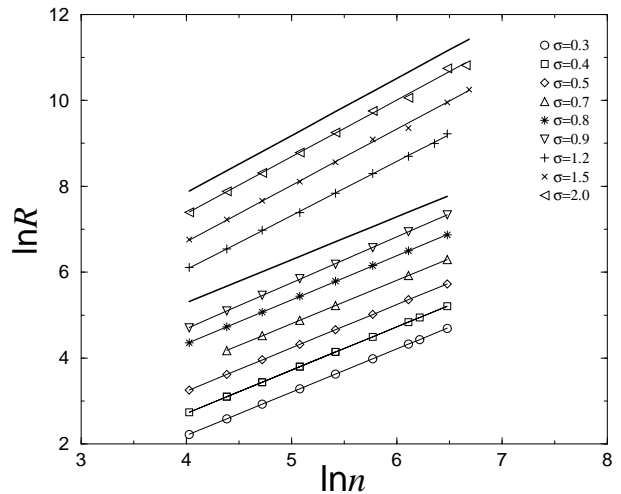,width=8cm}
\caption{Log-log plot of the average cluster size $R$ versus
the total number of particles $n$ for several values of $\sigma$
at the subcritical regime. The straight lines from top to
bottom have slopes 1.33704 and 1, respectively. The data points
have been shifted in order to avoid  overlapping.}
\label{fig4}
\end{figure}
In Fig. 4 we show the log-log plot of $R$ versus $n$ 
 in the  subcritical regime for 
several values of $\sigma$. For $\sigma>1$,
the slopes of the straight lines fitted to the data points 
are consistent with $1.33704(4)$, as expected
 for the DP universality class  whose 
clusters generated at the criticality \cite{vics92} have fractal dimension
 $d_{F}=0.74792(2)$ \cite{marr99}, 
whereas for $0<\sigma<1$ the slopes are equal to $1$. 
The change in the fractal dimension shows the change in the
nature of the phase transition, passing from  continuous to
discontinuous, as one decreases $\sigma$. A fractal
dimension equal to the Euclidean dimension is  actually the signature
of a first order transition, since we have a compact cluster that
remains finite in the thermodynamic limit.
To see this, let us evaluate the density $\rho^*$
of the cluster of particles $\rho^*=n/R$. It is worth emphasizing that
 although the   total density $\rho=n/L$ 
of the system is zero, since the system is constrained in the subcritical 
 regime, the quantity $\rho^*$ is finite. 
If the phase transition is first order, $\rho^*$ will have
a non-zero value in the thermodynamic limit. 
Rewriting  Eq. (\ref{fractal}) in terms of  $\rho^*$, 
one has the following relation $\rho^*=n/R \sim  n^{-(1-d_F)/d_F}$. 
Therefore, for $0<\sigma<1$, where $d_{F}=1$,
$\rho^* \rightarrow \rho^{*}_{0}$,  when $n
 \rightarrow \infty$, corresponding  to  the active phase
 $\rho^{*}_{0}$ in coexistence with the
absorbing phase ($\rho=0$). The density 
$\rho^{*}_{0}$ is determined simply by the inverse of the slope of
curves in Eq. (\ref{fractal}). 
In contrast, for a second
order transition in which $d_F<1$, $\rho^*$  vanishes when 
$n\to\infty$, because $\rho^*=n/R \sim n^{-(1-d_F)/d_F}\rightarrow 0$ when
$n \rightarrow \infty$.
The change in the nature of the phase transition for $\sigma$ close
to 1 is in agreement with results by Ginelli et al \cite{gine05} who
used time-dependent numerical simulations and  distribution of inactive
islands  to identify the order of the phase transition.

We close this section by comparing the density $\rho^{*}_{0}$ of the
compact cluster obtained from both ensembles. As it was mentioned
above, in the constant
particle number ensemble the density  $\rho^{*}_{0}$ is evaluated
directly, since it is simply the inverse of the slope of Eq. (\ref{fractal}).
To determine $\rho^{*}_{0}$ in the constant rate ensemble, 
we use the procedure 
adopted by Dickman \cite{dick91,maia07} 
that  consists of dividing the system into blocks of 100
sites and determining histograms of such blocks density profiles
at the phase coexistence. For $\sigma=0.5$ at the phase coexistence, the 
 histogram show a bimodal distribution with a peak at
$\rho^*=0$ and another  one at $\rho^{*}_{0}=0.77$. This agrees
  with the density of the compact cluster $\rho^{*}_{0}=0.777(1)$,
 obtained  from the present approach (constant particle
number ensemble). For other
values of $\sigma$ in the interval $0<\sigma<1$, the densities  
of  compact clusters at the phase coexistence  
obtained from both ensembles agree very well.

\subsection{Supercritical regime}

\begin{figure}
\epsfig{file=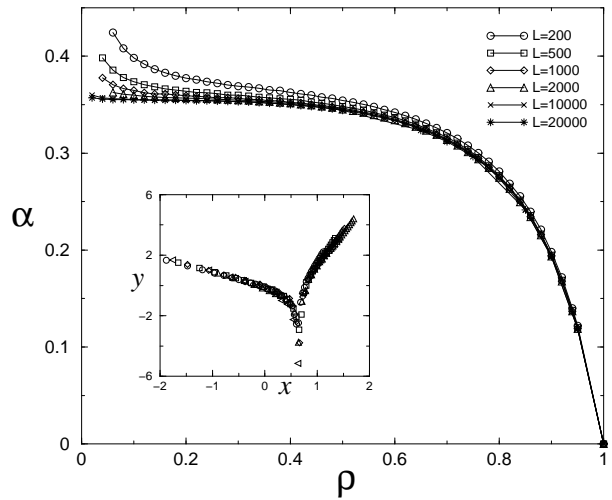,width=8cm}
\caption{Quantity $\alpha$ versus $\rho$ for several
values of the system size $L$ for $\sigma=2$. The inset
shows the collapse of the data by using relation (\ref{fsisup})
 and (\ref{fsisub})
where $y=L^{1/\nu_\perp}|\alpha-\alpha_0|$
and $x=L^{\beta/\nu_\perp}\rho$ for the supercritical regime
and $y=L^{d_F}|\alpha-\alpha_0|$
and $x=L^{d-d_F}\rho$
for the subcritical regime.}
\label{fig5}
\end{figure}

\begin{figure}
\epsfig{file=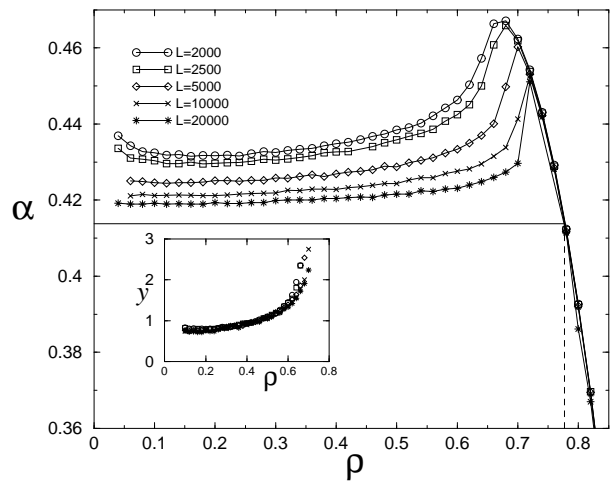,width=8cm}
\caption{Quantity $\alpha$ versus $\rho$ for several
values of the system size $L$ for $\sigma=0.5$. 
The horizontal straight line is the tie line at
$\alpha=\alpha_0=0.41389$. The dashed vertical line
indicates the density $\rho^*=0.777$ of the coexisting
active phase. The inset
shows the collapse of the data by using relation (\ref{asymf})
$y=(\alpha-\alpha_0)L^{\sigma}$.}
\label{fig6}
\end{figure}

The supercritical regime is characterized by  nonzero values
of the density $\rho$. To simulate the $\sigma-$contact process 
 in the supercritical regime, 
we have considered lattice sizes ranging from $L=200$ 
to $L=20000$. In Figs. 
\ref{fig5} and \ref{fig6} we have plotted the quantity $\alpha$ versus $\rho$
for several values of $L$ for $\sigma=2$ and $\sigma=0.5$,
respectively. The curves exhibit a strong dependence on $L$ that
is qualitatively different whether we consider $\sigma>1$ 
or $0<\sigma<1$. In the first case (exemplified in Fig. 
\ref{fig5} by $\sigma=2$),
the curves are strictly decreasing functions and cumulated, 
when $L\to\infty$, into a strictly decreasing function.

In the second case (exemplified in Fig. \ref{fig6} by $\sigma=0.5$),
the curves are no longer decreasing and present  ``loops'', 
that can be associated to a coexistence of two phases. As it was
shown previously,
the occurrence of ``loops'' in numerical simulations agrees
qualitatively  mean-field results. However, in contrast to
the mean-field approach, the existence of ``loops'' in numerical simulations
is  a finite size effect that  disappears when $L \rightarrow \infty$,
giving rise to a horizontal tie line that connects the two
coexisting phases.          

For $\sigma>1$, where the phase transition is second order,  the finite
size scaling  for the density is given by \cite{tome01}
\begin{equation}
\label{fsisup}
\rho = L^{-\beta/\nu_\perp} f (\epsilon L^{1/\nu_\perp}),
\end{equation}
valid for the supercritical regime,
where $\epsilon=\alpha-\alpha_0$ and $\beta$ and $\nu_\perp$
are critical exponents associated to the order parameter and
the spatial correlation length, respectively; and  \cite{tome01}
\begin{equation}
\label{fsisub}
\rho = L^{-d+d_F} f (\epsilon L^{d_F}),
\end{equation}
valid in the subcritical regime,
where $f(x)$ is an universal function. 
A data collapse obtained by
using the finite size scalings (\ref{fsisup}) and (\ref{fsisub}) 
and the best estimates of the DP critical exponents
$\beta=0.276486(8)$, $\nu_{\perp}=1.096854(4)$ and $d_{F}=0.74792(2)$
\cite{marr99} is shown in the inset of Fig. \ref{fig5}.
\begin{figure}
\epsfig{file=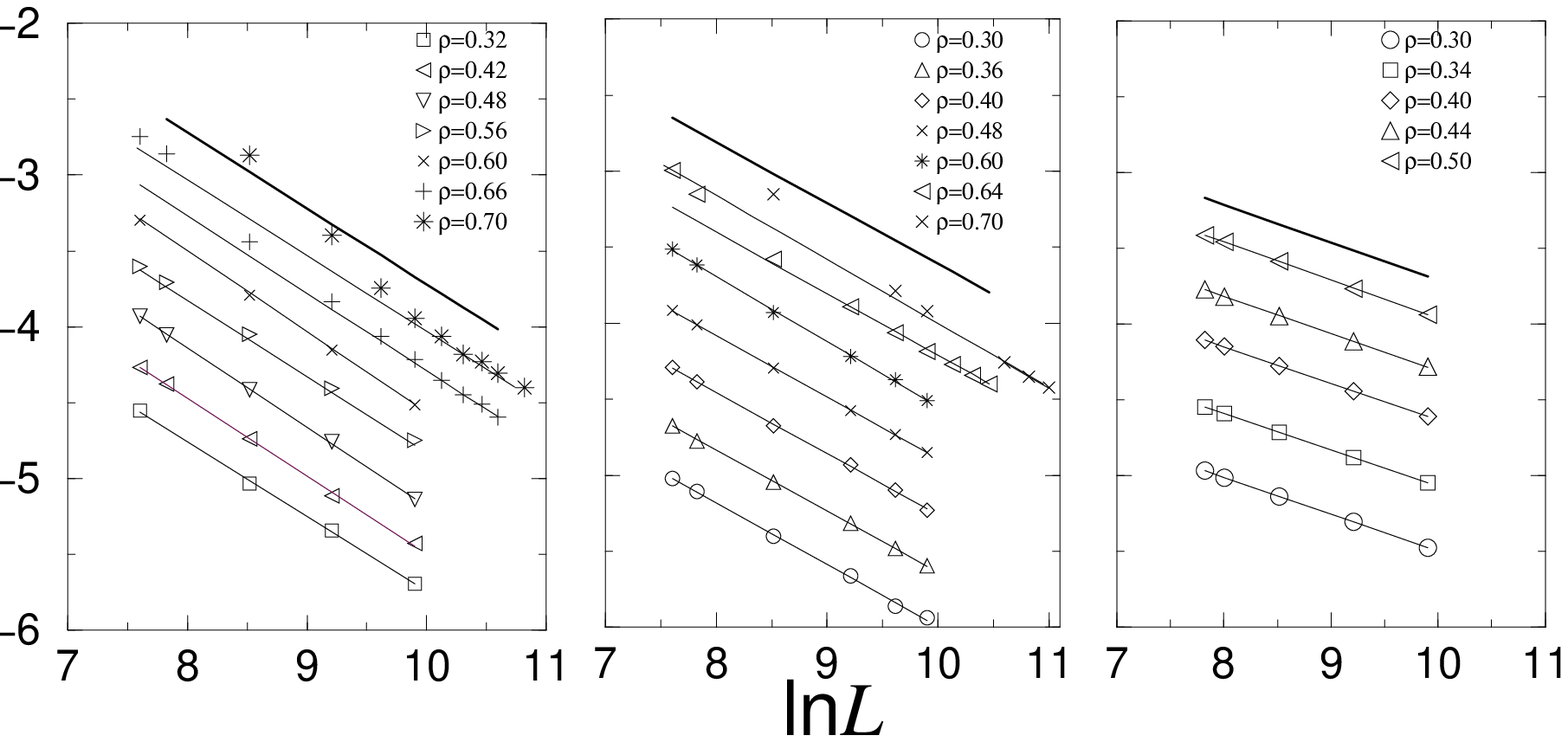,width=10.cm,height=6.5cm}
\caption{Log-log plot of $\alpha-\alpha_0$ versus $L$ for several densities
$\rho$ for $\sigma=0.5$ (left panel),  $\sigma=0.4$ (center panel) 
and $\sigma=0.25$ (right panel). From left to right, the straight lines at each
panel have slopes $0.5$, $0.4$ and $0.25$. The data points
have been shifted in order to avoid  overlapping.}
\label{fig7}
\end{figure}

On the other hand, 
for $0<\sigma <1$, where the phase transition is first order, Eqs. 
(\ref{fsisup}) and (\ref{fsisub}) 
are not valid. To relate the parameter $\alpha$ with the system size $L$, 
we assume the following behavior of $\alpha$ in the region of the
phase coexistence
\begin{equation}
\label{asymf}
\alpha - \alpha_0 \sim L^{-\sigma},
\end{equation}
where $\alpha_0$ is the value of $\alpha$ at the
coexistence, or the value of $\alpha$ where the
tie line is located. 

An argumentation concerning Eq. (\ref{asymf}) is given
by assuming the one-site mean-field approximation given by
Eq. (\ref{eqmf}). For $0<\sigma<1$,
the sum on the right side of Eq. (\ref{eqmf}) can be replaced by an
integral, which to leading order in $\rho$, reduces to
\cite{gine05,gine06}
\begin{equation}
\alpha-\bar{\alpha_{0}}=-\rho+a\Gamma (1-\sigma)\rho^{\sigma}.
\label{eq13}
\end{equation}
where $\bar{\alpha_{0}}=1$ is its one-site mean-field transition value
and $\Gamma (x)$ is the gamma function.
Relating $\rho$ with the system size $L$,
we have that $\alpha-\bar{\alpha_{0}}\sim L^{-\sigma}$, because
the second term in the right-hand side of Eq. (\ref{eq13}) dominates
over the first.

The correctness of Eq. (\ref{asymf}) can be checked by the log-log plot of
$\alpha - \alpha_0$ versus $L$ by using the
estimation of $\alpha_0$ available from the 
study of the subcritical regime. Indeed,
as shown in Fig. \ref{fig7} the slopes of the straight
lines fitted to the data points for several
densities are consistent with the values of $\sigma$.
 We have repeated  the analysis   for other values
of $\sigma$ in the interval $0<\sigma<1$, whose
dependence  of $\alpha$ on the system size $L$
at the phase coexistence is also described by Eq. (\ref{asymf}). 
For  higher densities close to $\rho^*$, in which one
observes a peak of $\alpha$ versus $\rho$, it is necessary
to consider larger system sizes in order to reach the asymptotic
behavior described by Eq. (\ref{asymf}), as can be seen for
the highest densities 
in the left and center panels  of Fig. \ref{fig7}, respectively.

\begin{figure}
\epsfig{file=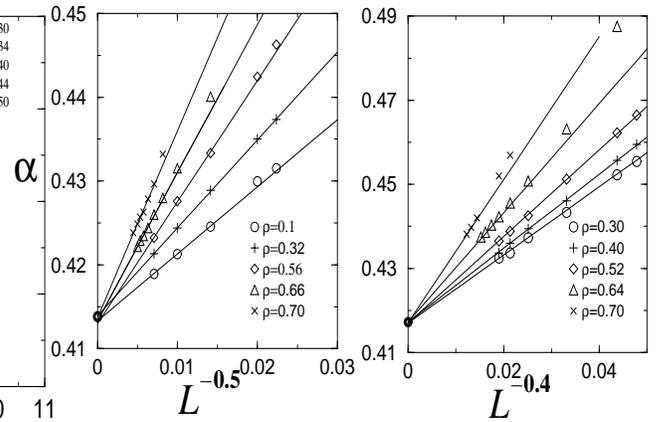,width=8.5cm,height=6cm}
\caption{Dependence of the parameter $\alpha$ on  $L^{-\sigma}$ 
for $\sigma=0.5$ (left) and $\sigma=0.4$ (right) 
considering several densities $\rho$.
The  full circles are  extrapolated values of $\alpha_0$ obtained
from the subcritical regime.}
\label{fig8}
\end{figure}

A complementar analysis, but fully equivalent,
consists in assuming Eq. (\ref{asymf}) and using it 
for determining estimates of $\alpha_0$ by numerical extrapolation,
 as shown in Fig. \ref{fig8}.
The agreement between extrapolated $\alpha_0$ using Eq. (\ref{asymf})
and those estimates of $\alpha_0$ available from the subcritical regime 
and spreading simulations (reported in the next section) confirms the
equivalence of ensembles in the thermodynamic limit at the phase coexistence.

As a final checking, we plot in the inset of
Fig. \ref{fig6} a collapse of the data for different
system sizes by considering the variable 
$y=(\alpha-\alpha_0)L^\sigma$ versus $\rho$. This confirms once more
the conjecture given by Eq. (\ref{asymf}).

\subsection{Time dependent numerical simulations}\label{time}
Here, we show explicitly
 results from time-dependent numerical simulations
for the $\sigma-$contact process considering $\sigma=0.4$, in
 order to compare the results for the estimation of $\alpha_0$ from
distinct ensembles.
 
Starting from a configuration close to the absorbing state, 
this procedure consists in determining 
 the time evolution of appropriate quantities, such
as the survival probability $P_s(t)$, the total number of particles
$N_p(t)$ and the mean square spreading $R^2(t)$ of the active region.
At the emergence of a second-order transition, these quantities are
described by following power-law behaviors 
\begin{equation}
P_s(t)\sim t^{-\delta}, \quad N_{p}(t)\sim t^{\eta} \quad  {\rm and} \quad
R^2(t)\sim t^{z},
\label{eq16}
\end{equation}
where $\delta,\eta$ and $z$ are their associated critical
exponents. They are related to the fractal dimension introduced in
 Eq. (\ref{fractal}) through the relation $d_{F}=2(\eta+\delta)/z$.

Although the order parameter presents
a jump in a nonequilibrium 
first-order transition, some dynamic variables are also  
characterized by dynamic exponents \cite{odor04}. In the
present case, the quantities  from Eq. (\ref{eq16}) 
are expected to be described
by the same exponents as the Glauber-Ising model at zero temperature, which
values are $\delta=1/2$, $\eta=0$ and $z=1$ \cite{gine05}.  Relating
these exponents, we have that $d_F=1$, in agreement with results
obtained previously from analysis in the subcritical regime for $0<\sigma<1$.

\begin{figure}
\vspace{1cm}
\epsfig{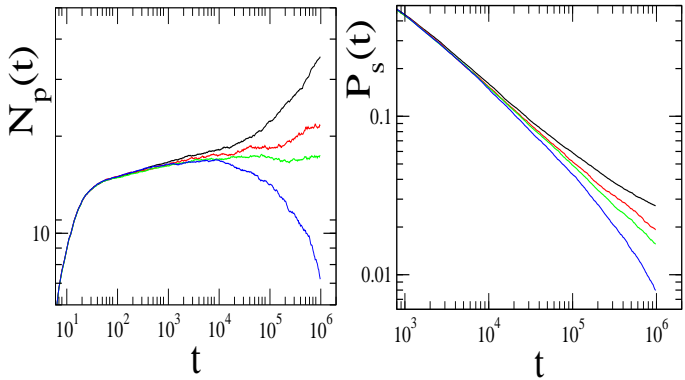}
\caption{Temporal evolution of the total number of particles 
$N_p$ (right) and the survival probability $P_s$ (left) for the ordinary
$\sigma-$contact process for $\sigma=0.4$. From top to bottom,
we have $\alpha=0.4167$, $0.4171$, $0.4172$ and $0.4176$.}
\label{fig9}
\end{figure}

In Fig. (\ref{fig9}), we show the plot of temporal evolution
of the quantities $P_s$, $N_p$  for some values of $\alpha$. 
For $\alpha=0.4172$, 
the quantities $N_p(t)$ and $P_s(t)$ follow indeed a power-law behavior
described by Eq. (\ref{eq16}), whose exponents are consistent to 
$\eta=0$ and $\delta=1/2$, respectively. 
This estimation of $\alpha_0$ agrees very well to those obtained from the
constant particle number ensemble, confirming the equivalence of ensembles.

\section{Conclusion}

In this paper we have studied a nonequilibrium model  
with long-range interactions, named $\sigma-$contact process,
in the ensemble of constant particle number. 
The equivalence of  ensembles 
is confirmed by the excellent agreement between the numerical 
results coming from both ensembles. 
All results obtained by the present approach are in full
agreement with those obtained by Ginelli et al \cite{gine05}.
We believe that the present
approach may be particularly useful for studying discontinuous 
phase transition, since it is possible to identify the nature
of the transition by measuring the spatial structure of
particles at the transition. The interesting feature
of the present approach concerns the study of  an infinite
system with a finite number of particles. As the number
of particles is increased, the systems naturally approaches
the critical point and, in this sense, it behaves like
self-organized critical systems.

Another advantage of the present approach is verified when one
 studies the  system in the supercritical regime. 
The dependence on the system size
is distinct in both regimes as shown in Figs. \ref{fig5} and \ref{fig6}.
The presence of a ``loop'' is a strong indication of
a first-order transition. Of course, results coming from finite
systems, such as the ``loop'' presented in Fig. \ref{fig6} is a particularity
of the conserved ensemble, that disappears in the thermodynamic limit.

We remark that the use of this
procedure in the constant rate ensemble is not possible
since numerical simulations of finite systems will present
a jump in the order parameter close to the transition point,
even in the case of a second-order transition, due
to the presence of the absorbing state.
Recently, de Oliveira and Dickman \cite{dic05} have proposed a method,
 named quasi-stationary simulations,  that improves the accuracy of
 results, even at the emergence of the phase transition. It consists in
 simulating the system in the constant rate ensemble in the standard
 way.
However, whenever the system enters in the absorbing state, a
 non-absorbing configuration is chosen from 
a list of  saved periodically configurations and then system returns to
 the active state. This method has  revealed an useful
tool in the study of systems  with absorbing states \cite{gine06,maia07,dic06}.
Other techniques, such as hysteretic analysis in a constant
coverage ensemble have also been used  to study discontinuous
 transition in nonequilibrium systems \cite{ziff86,losc02,mone01}.
 These approaches are, however, distinct from ours,
in the sense that the standard ensemble is initially used to generate
a stationary configuration and then the system is switched to the
constant coverage
ensemble.

\section*{ACKNOWLEDGEMENT}
C. E. F. acknowledges the financial support from    
Funda\c c\~ao de Amparo \`a Pesquisa do   
Estado de S\~ao Paulo (FAPESP) under Grant No. 06/51286-8.


\begin{thebibliography}{99}
\bibitem{loops} Loop is an expression 
borrowed from  equilibrium 
statistical mechanics  to design an region of instability, in analogy
with the van-der-Waals loops. In this case,
the stability conditions are obtained by
performing the Maxwell construction.
\end{thebibliography}
\end{document}